\documentclass{article} 
\usepackage{iclr2026_conference}
\usepackage{times}

\usepackage{amsmath,amsfonts,bm}

\def\eqref#1{equation~\ref{#1}}

\def\1{\bm{1}}

\DeclareMathAlphabet{\mathsfit}{\encodingdefault}{\sfdefault}{m}{sl}
\SetMathAlphabet{\mathsfit}{bold}{\encodingdefault}{\sfdefault}{bx}{n}

\usepackage{hyperref}
\usepackage{url}
\usepackage{graphicx}
\usepackage{booktabs}
\usepackage{listings}

\setcitestyle{numbers,square,sort&compress}

\title{s2n-bignum-bench: A practical benchmark for evaluating low-level code reasoning of LLMs}

\author{%
Balaji Rao \thanks{Extended work done after completing internship at AWS; Corresponding author} \\
  Stevens Institute of Technology\\
  Hoboken, NJ 07030 \\
  \texttt{brao@stevens.edu} \\
   \And
John Harrison \thanks{Work independent from role at AWS}\\
    Amazon Web Services \\
    Seattle, WA 98101 \\
   \texttt{jargh@amazon.com} \\
   \And
Soonho Kong \footnotemark[2] \\
   Amazon Web Services\\
   Seattle, WA 98101  \\
   \texttt{soonho@amazon.com} \\
   \And
Juneyoung Lee \footnotemark[2] \\
    Amazon Web Services \\
    Seattle, WA 98101 \\
   \texttt{lebjuney@amazon.com} \\
   \And
Carlo Lipizzi \\
  Stevens Institute of Technology \\
  Hoboken, NJ 07030\\
   \texttt{clipizzi@stevens.edu} \\
}

\iclrfinalcopy 
\begin{document}

\maketitle

\begin{abstract}
Neurosymbolic approaches leveraging Large Language Models (LLMs) with formal methods have recently achieved strong results on mathematics-oriented theorem-proving benchmarks. However, success on competition-style mathematics does not by itself demonstrate the ability to construct proofs about real-world implementations. We address this gap with a benchmark derived from an industrial cryptographic library whose assembly routines are already verified in HOL Light. s2n-bignum is a library used at AWS for providing fast assembly routines for cryptography, and its correctness is established by formal verification. The task of formally verifying this library has been a significant achievement for the Automated Reasoning Group. It involved two tasks: (1) precisely specifying the correct behavior of a program as a mathematical proposition, and (2) proving that the proposition is correct. In the case of s2n-bignum, both tasks were carried out by human experts. In \textit{s2n-bignum-bench}, we provide the formal specification and ask the LLM to generate a proof script that is accepted by HOL Light within a fixed proof-check timeout. To our knowledge, \textit{s2n-bignum-bench} is the first public benchmark focused on machine-checkable proof synthesis for industrial low-level cryptographic assembly routines in HOL Light. This benchmark provides a challenging and practically relevant testbed for evaluating LLM-based theorem proving beyond competition mathematics. The code to set up and use the benchmark is available here: \href{https://github.com/kings-crown/s2n-bignum-bench}{s2n-bignum-bench}.
\end{abstract}

\section{Introduction}

Formal theorem proving with Large Language Models (LLMs) and interactive theorem provers has become a central testbed for LLM reasoning, but existing benchmarks  emphasize competition-style mathematical problems. Solving complex math problems requires a rigorous framework of steps and logical proofs, and success on such tasks evidences structured reasoning~\cite{achim2023harmonic}. However, excellence on math-centric benchmarks does not automatically transfer to systems with practical engineering consequences. Therefore, the design of diverse and high-quality benchmarks is a key challenge in this research area.

To complement existing benchmarks, we propose \textit{s2n-bignum-bench}, a machine-checkable benchmark distilled from the \textit{s2n-bignum} cryptographic library, focusing on verified low-level code. The benchmark tests whether LLMs can synthesize machine-checkable proofs about real low-level implementations rather than only competition-style mathematics.

Our contributions are fourfold. First, we package \textbf{2{,}284}\footnote{As of this submission; extracted from \textit{s2n-bignum}. V1.0, pinned to commit \texttt{9912d17...} at \href{https://github.com/kings-crown/s2n-bignum}{s2n-bignum}} proof obligations from the production-grade \textit{s2n-bignum} cryptography library as isolated \emph{context--query} tasks with stable per-problem identifiers and standalone artifacts. Second, we provide an end-to-end pipeline to build the benchmark, retrieve selected problems, and run fully offline evaluation using the shipped artifacts. Third, we include integrity mechanisms that detect unsound or invalid submissions, including checks for newly introduced axioms, forbidden placeholders such as \texttt{CHEAT\_TAC}, and parser-level validation of submitted proof expressions. We also provide a contamination-mitigation mechanism based on type-annotation obfuscation. Fourth, because the benchmark is grounded in a deployed cryptographic codebase, it measures ISA-aware, bit-precise reasoning that is closer to real verification workflows than competition mathematics.

\section{Background and Related Work}

\subsection{Theorem Proving}

Interactive theorem provers (ITPs) following the LCF approach all have their derivations ultimately checked by a
small, trusted kernel that produces values of type \texttt{thm} (the “theorem” type)~\cite{harrison2014history}. Examples of ITPs include HOL Light, Lean4, Isabelle/HOL, and Rocq Prover.  HOL Light is a minimalist proof system designed for higher-order logic (HOL) implemented in OCaml, with a very small trusted kernel and an emphasis on clarity~\cite{harrison2009hol}. Its proof script is an OCaml program and the proof system  uses the OCaml toplevel (REPL) for interactivity. It relies on tactics to discharge goals.

LLMs have become a central part of \emph{neural theorem proving} (NTP), where the model proposes proof steps while an interactive theorem prover (ITP) acts as the verifier. Unlike answer-only benchmarks (e.g., gsm8k~\cite{cobbe2021training}, CruxEval~\cite{gu2024cruxeval}), NTPs demand \emph{structured, verifiable reasoning}.

\subsection{Theorem Proving Benchmarks}

MiniF2F is the most widely adopted cross-system benchmark, with 488 formalized Olympiad-level mathematics problems translated across multiple proof systems including Lean4, Metamath, Isabelle, and HOL Light~\cite{zheng2022minif2fcrosssystembenchmarkformal}, saturated in 2025 by Seed Prover~\cite{chen2025seedproverdeepbroadreasoning}. PutnamBench introduces competition mathematics from the William Lowell Putnam Mathematical Competition, featuring 1,724 hand-constructed formalizations of 672 theorems across Lean4, Isabelle, and Rocq~\cite{tsoukalas2024putnambenchevaluatingneuraltheoremprovers}, 99.4\% solve rate by Aleph~\cite{Aleph}. 

Recent benchmarks have expanded toward verification conditions and repository-scale software verification. In particular, NTP4VC studies theorem proving over verification conditions extracted from real systems code, while VeriSoftBench evaluates proof synthesis over repository-scale Lean verification tasks\cite{xu2026neural, xin2026verisoftbench}. miniCTX and VeriBench-FTP are designed to test the use of context as well as theorem-level, context-level, and project-level generalization across several mathematical, as well as code domains~\cite{hu2024minictx, barkallah2025veribench}. These works represent a paradigm shift toward realistic theorem proving scenarios where models must leverage extensive context from real Lean projects. Other relevant works that have moved toward verification-oriented proving and code-centered formal reasoning, include miniCodeProps, CLEVER, and VERINA~\cite{lohn2024minicodeprops, thakur2025clever, ye2025verina}. Works like SorryDB emphasize the need for dynamically-updating benchmarks~\cite{letson2026sorrydb}. Our work is complementary to these efforts: we focus specifically on HOL Light proof synthesis for industrial low-level cryptographic assembly with shipped object-code artifacts and trusted ISA semantics.

\section{Motivation}

Recent work has extended neural theorem-proving evaluation beyond competition mathematics toward software verification and repository-scale proof synthesis. We aim to supplement these works by proposing an underrepresented ecosystem: mechanized proofs about industrial low-level cryptographic assembly in HOL Light. While some benchmarks have been effective at evaluating reasoning models, they do not test whether models can construct machine-checkable proofs about real implementations. The \textit{s2n-bignum} proofs require a form of reasoning that is qualitatively distinct from both abstract mathematics and higher-level verification-condition proving. Each proof must show that starting from a precondition on registers and memory, a specific sequence of decoded ARM or x86 instructions produces a final state satisfying a mathematical postcondition. Proving this involves decomposing the program at specific program-counter offsets, symbolically executing each segment by rewriting through ISA-specific decode and execute semantics, and simplifying the resulting symbolic state terms at each step. Math-centric proving does not generally involve architectural state, aliasing, or endianness, and competence in abstract mathematics does not, by itself, establish capability for low-level code reasoning.

The correctness of cryptographic libraries and systems code has immediate security and reliability consequences; this style of low-level implementation reasoning remains underrepresented in theorem-proving benchmarks. The s2n-bignum library contains hand-tuned big-integer assembly subroutines (x86/ARM) accompanied by HOL Light proofs that the object code meets functional correctness  specifications under a trusted ISA model. Building a benchmark from this corpus lets us evaluate an NTP system's ability to construct a proof that real assembly satisfies its specification.

\begin{table}[h!]
\centering
\small
\resizebox{\linewidth}{!}{%
\begin{tabular}{lllllll}
\toprule
\textbf{Benchmark} 
& \textbf{Formal system} 
& \textbf{Primary focus} 
& \textbf{Problems (\#)} 
& \textbf{Task setting}\\

\midrule
miniF2F~\cite{zheng2022minif2fcrosssystembenchmarkformal} 
& Multiple
& Olympiad mathematics (formal) 
& 488 
& Formal theorem proving\\

PutnamBench~\cite{tsoukalas2024putnambenchevaluatingneuraltheoremprovers} 
& Multiple 
& Competition mathematics (formal) 
& 672 
& Formal theorem proving\\

NTP4VC~\cite{xu2026neural}
& Multiple
& Verification conditions from real code 
& 600 
&  Real-world VC proving\\

VeriBench-FTP~\cite{barkallah2025veribench}
& Lean4
& Code-verification artifacts
& 857
& Proofs from verification artifacts\\

miniCTX/miniCTX-v2~\cite{hu2024minictx} 
& Lean4 
& Context-dependent proving  
& 762 
& Context / Project generalization\\

VeriSoftBench~\cite{xin2026verisoftbench}
& Lean4 
& Repository-scale software verification 
& 500 
& Repository-scale verification proving\\

\midrule
\textbf{s2n-bignum-bench} & HOL Light & Verified cryptographic assembly programs
& \textbf{2284} 
& Proof synthesis over machine-code\\

\bottomrule
\end{tabular}
}
\caption{\textit{s2n-bignum-bench} relative to representative theorem-proving and verification benchmarks}
\label{tab:unified-benchmark}
\end{table}
\vspace{-1.2em}

\section{s2n-bignum-bench Construction}
Our problems are derived from the open-source \textit{s2n-bignum} repository, an AWS cryptographic library. Each problem in \textit{s2n-bignum-bench} is a HOL Light \emph{context--query} task. We inline the relevant OCaml modules, locate top-level theorem bindings of the form \texttt{let THM = prove(goal, proof)}, and extract the goal as the \emph{query}. The accompanying \emph{context} is a self-contained OCaml/HOL Light setup that loads the required definitions, constants, and previously proved results needed to reproduce the original proving environment, while replacing each original proof body with the placeholder \texttt{CHEAT\_TAC}.\footnote{\texttt{CHEAT\_TAC} is a placeholder tactic in HOL Light, analogous to \textit{sorry} in Lean or Isabelle. Any submission that uses it is rejected as cheating.} With this, we isolate the task of synthesizing a new, machine-checkable proof under the same interfaces and imports as the source project.

To distinguish between different problems, we introduce the notion of a \texttt{Problem identifier}, since the same theorem name may appear across different proof files or multiple times within the same file. A problem identifier has the form \texttt{{arch}.{filename}.{thm}.{N}}, for example: \texttt{arm.bignum\_montsqr\_p256.lemma1.0}, where $N$ represents an occurrence index of a lemma. 
In this work, we also include the artifacts to extract selected problems into a directory with (i) \texttt{setup.ml} and (ii) \texttt{query.txt}. This will allow for reproducible, standalone attempts per problem.

Using lightweight heuristics over theorem names and goal forms, we partition the benchmark into four categories: \texttt{Bit-vector lemmas} (\textbf{311}), \texttt{Program-state lemmas} (\textbf{552}), \texttt{Functional correctness} (\textbf{859}, comprising \textbf{437} ARM and \textbf{422} x86 problems), and \texttt{Generic} (\textbf{562}) for auxiliary facts not captured by the preceding categories.

HOL Light proofs are typically developed interactively through the OCaml REPL. Existing tooling, such as \texttt{hol\_server} and the VSCode extension for HOL Light, can be used to provide an interactive development environment on top of the released benchmark artifacts~\cite{HOLVC}.

\section{Evaluation}

\subsection{Answer submission and grading}

Challengers submit a proof expression along with the name of the problem attempted. We first perform a syntax and type pre-check by compiling a generated \texttt{.synchk.ml} file that pastes the submitted proof expression into the benchmark context. This catches malformed tactic expressions and other immediate parser or type errors before full evaluation.

Each submitted proof attempt yields exactly one verdict per problem: \texttt{OK}, \texttt{FAIL}, \texttt{CHEATING}, \texttt{TIMEOUT}, or \texttt{ERROR}. Results are aggregated into a CSV file, and the primary task metric is binary success at kernel-checked proof completion. To make model comparisons meaningful, an \emph{official} evaluation configuration should fix the proof-check timeout, hardware setting, and submission budget. We also provide support for user-configurable timeouts for exploratory use\footnote{\label{fn:timeout}Detailed explanation in Appendix}. Auxiliary lemmas may be defined inside the submitted proof expression, provided that the overall expression evaluates to a tactic.

The model is not given access to the original proof bodies or the tactics used to prove other theorems. This restriction is important for preserving the validity of the benchmark. It is important to note, however, that the challenger can provide the relevant machine-code context needed to understand the specification being proved to the LLM.

\subsection{Initial Baseline Experiments}
As a preliminary baseline, we evaluate GPT-5.3-Codex~\cite{5.3-Codex} through \texttt{codex-cli} under the configuration described in Appendix~A. The model achieves a binary proof-completion rate of \textbf{4.4\%} in medium-effort mode and \textbf{5.3\%} in high-effort mode over the full benchmark. We treat this as an initial baseline rather than an exhaustive estimate of current model capability\textsuperscript{\ref{fn:timeout}}.

\subsection{Integrity and contamination defenses}

To mitigate contamination from memorized theorem statements, we implement an obfuscation mechanism that makes type annotations more explicit. We set the \texttt{print\_types\_of\_subterms} of HOL Light to the most verbose mode and reprint the queries. However, this works for only $\approx$70\% of the problem set because HOL Light's printer and parser are not fully Pollack-consistent ~\cite{wiedijk2012pollack}. For such queries, we chose to use their original representations without obfuscation\footnote{We are communicating with HOL Light's maintainers to fix this}.

To ensure that a challenger did not introduce any forbidden tactics like \texttt{CHEAT\_TAC} or functions like \texttt{new\_axiom}, our evaluation checks the output of the \texttt{axioms()} function in HOL Light. The function returns a list of theorems that have been axiomatized so far. If the answer from challenger did not use any forbidden functions, the result of \texttt{axioms()} must be identical before and after the solution. If they are different, our benchmark script marks the result as ``CHEATING''.

A separate class of attacks attempts to introduce a syntactically complex solution that resembles ``SQL injection''. To prevent this, we invoke an OCaml parser for each solution and check whether it has one valid expression. Submissions that fail this check are rejected before proof evaluation.

\section{Conclusion}

We introduce \textit{s2n-bignum-bench}, a benchmark for machine-checkable proof synthesis over a deployed corpus of verified low-level cryptographic assembly proofs in HOL Light. This benchmark targets a capability that remains underrepresented in current evaluation: constructing sound proofs about real low-level implementations under trusted ISA semantics. By releasing isolated problem artifacts, an offline evaluation harness, and integrity checks against unsound submissions, we aim to provide a reproducible testbed for future work on theorem proving and verification-oriented reasoning beyond competition mathematics. Although the current release focuses on functional correctness, recent HOL Light developments around \textit{s2n-bignum} also formalize relational properties, including constant-time discipline and equivalence between optimized and verification-friendly routines, suggesting a natural path toward future benchmark extensions beyond extensional correctness~\cite{mazzucato2025relational,s2nSound}.

\newpage
\bibliographystyle{chicago}
\bibliography{iclr2026_conference}

\newpage
\appendix

\section{A guide for challengers}
The official repository walks a user through setting up the benchmark, and the experimental protocol overview described here is a good place to start to begin exploring this problem set.

\subsection{Experimental protocol overview}
We conducted preliminary \emph{pass@1} experiments to evaluate whether current language models can synthesize HOL Light tactic proofs for \textit{s2n-bignum-bench}. The experiment(s) follow the benchmark workflow described in the repository documentation: \href{https://github.com/kings-crown/s2n-bignum-bench}{s2n-bignum-bench}.

Our primary reported baseline uses a closed-source reasoning model accessed through the Codex-CLI. The zero-shot prompt template and evaluation scripts are available in official repository.

\subsection{Benchmark preparation and problem retrieval}

Following the repository workflow, we first build the benchmark and extract theorem metadata from the pinned \textit{HOL Light} and \textit{s2n-bignum} sources. This produces a corpus of \textbf{2,284} problems. Each problem consists of:
\begin{itemize}
    \item a HOL Light boolean term stored in \texttt{query.txt}, and
    \item a corresponding \texttt{setup.ml} file containing the HOL Light session preamble needed to establish the proof context.
\end{itemize}

For inference, problems were (and can be) retrieved in two equivalent formats:
\begin{itemize}
    \item as a flat CSV using \texttt{retrieve-problem.py ....... --csv-only}, and
    \item as a directory tree of \texttt{<problem-id>/query.txt} files under a problems directory.
\end{itemize}

\subsection{Prompting and inference pipelines}
\paragraph{Prompt template.} All runs use the same zero-shot prompt template, using the repository prompt:

\begin{quote}\small
\begin{verbatim}
You are an expert in HOL Light. I am going to give a HOL 
Light boolean term. Please write a HOL Light proof of it 
in a THEN form. Do not use CHEAT_TAC or new_axiom.

<Example>

If the input is

`x * (y + z) = x * z + x * y`

A possible answer is

REWRITE_TAC[LEFT_ADD_DISTRIB] THEN
GEN_REWRITE_TAC LAND_CONV [ADD_SYM] THEN
REFL_TAC

Please include the proof in your proof but not other natural 
statements, so that I can easily evaluate your answer.

<Query>

{query.txt}
\end{verbatim}
\end{quote}

The prompt includes a single worked example and instructs the model to output only a tactic expression, with no surrounding explanation. This matches the expected input format of the evaluation pipeline.

At this point, we have built the benchmark, and we have all the required components to evaluate the answers that are generated by the LLM.

\subsection{Example problem}
\texttt{x86.sha3\_keccak\_f1600.WORD\_NEG\_EL\_DEMORGAN} from the benchmark asks the prover to establish a De Morgan identity over machine words:
\begin{small}
\begin{verbatim}
`!(p:N word) (q:N word).
    (word_or p (word_not q)) = 
      word_not(word_and (word_not p) q)`
\end{verbatim}
\end{small}

The ground-truth proof discharges this in two tactics: 
\begin{small}
\begin{verbatim}
REPEAT GEN_TAC THEN WORD_BITWISE_TAC
\end{verbatim}
\end{small}

An alternative accepted proof does it through re-writes:
\begin{small}
\begin{verbatim}
REWRITE_TAC[WORD_NOT_AND] THEN
REWRITE_TAC[WORD_NOT_NOT] THEN
REFL_TAC
\end{verbatim}
\end{small}
This illustrates that multiple distinct tactic sequences can be used to solve the same goal. And our benchmark simply accepts any \texttt{correct} proof expression.

\subsection{Codex CLI baseline.}
Our main preliminary baseline uses GPT-5.3-Codex through \texttt{codex-cli}. For each problem, we substitute the goal term into the prompt and invoke the model in a restricted read-only sandbox with shell access, databases, and web search disabled. Responses are written to a CSV together with problem identifiers and auxiliary metadata. We also parse the CLI event stream to detect unexpected tool-related events; these are logged for audit purposes but are not used for evaluation.

\subsection{Construction of the problem timeout mapping}
The benchmark evaluator executes each submitted proof attempt with a timeout in order to prevent runaway tactics from consuming unbounded compute. We support two timeout mechanisms: category-level defaults in \texttt{timeouts.json} and per-problem overrides in \texttt{timeout-map.json}. In practice, proof-check times in \textit{s2n-bignum-bench} vary by several orders of magnitude, from a few seconds or milliseconds for small HOL Light lemmas to multiple hours for the largest routine-correctness theorems. 

A single blanket timeout therefore creates an undesirable tradeoff: if set too low, it incorrectly penalizes legitimate but expensive proofs; if set too high, failed submissions may waste hours before timing out, this might also lead to OOM issues which could lead to other problems.

To address this, we supply a heuristically derived per-problem timeout map based on repeated profiling of the ground-truth proofs.

To construct this map, we profile the original human-written proofs using the same benchmark evaluation harness that is later used for submitted answers. During profiling, category-level timeouts are temporarily set to a very large value so that the ground-truth proofs are not prematurely terminated. The evaluator then assembles, compiles, and executes the benchmark proof files exactly as in normal assessment, while recording per-problem wall-clock time for the internal \texttt{prove(goal, tactic)} call. We refer to this measurement as \texttt{prove\_secs}; it excludes compilation overhead and captures only proof execution time inside HOL Light.

Because proof execution times vary across runs due to operating-system scheduling, garbage collection, memory pressure, and parallel execution effects, we repeat this profiling procedure three times over the full benchmark. Across all runs, every ground-truth proof completes successfully. The repeated runs reveal substantial variance for some heavy proofs, including large absolute spreads for the most expensive routine-correctness theorems.

For each problem, we collect the profiled \texttt{prove\_secs} values from all profiling runs and compute summary statistics including the maximum runtime and an empirical high-percentile runtime. We then divide problems into two broad classes:
\begin{itemize}
\item \textbf{Light problems}: proofs whose profiled runtimes remain comfortably below the heavy-proof regime.
\item \textbf{Heavy problems}: proofs whose observed runtimes indicate substantially higher cost or variance.
\end{itemize}

Light and heavy problems are assigned timeouts using different multiplicative safety margins. Intuitively, heavy problems receive more generous scaling because they exhibit larger absolute runtime variance. In both cases, the assigned timeout is bounded below by a fixed minimum floor (120 seconds) and above by a global cap (10,800 seconds). This yields a timeout map that is conservative enough to accommodate runtime variation while still avoiding the extreme inefficiency of a blanket timeout.

The resulting \texttt{timeout-map.json} contains one timeout entry per benchmark problem. During evaluation, the benchmark first checks whether a submitted problem identifier appears in this map. If so, the corresponding per-problem timeout is used. If not, the evaluator falls back to the category-level default from \texttt{timeouts.json}. This fallback is primarily intended for exploratory use or for newly added benchmark problems that have not yet been profiled.

\subsection{Evaluation pipeline}
The inference pipeline is left up to the challenger to iterate on; with their own models or with API end-point based prompting with more intricate techniques. But ultimately for evaluation, the same artifact is expected: a directory of \texttt{<problem-id>/answer.txt} files. 

Evaluation proceeds in three stages:
\paragraph{(1) Syntax checking and assembly.}
Each candidate answer is first validated against the benchmark problem identifiers and then syntax-checked by compiling it in context as a HOL Light tactic expression. Concretely, the answer is wrapped into a small OCaml/HOL Light scaffold and checked with the benchmark syntax-checking script. Problems that fail this stage are excluded from further execution. 

\paragraph{(2) Proof execution.}
Answers that pass syntax checking are assembled into benchmark evaluation files and executed in HOL Light using the repository evaluation harness. Each proof attempt is run with:
\begin{itemize}
    \item a per-problem timeout, 
    \item pre/post axiom-count comparison to detect unsound submissions, and
    \item standard compile-and-run logging.
\end{itemize}

\paragraph{(3) Verdict collection.}
Each problem receives exactly one verdict:
\texttt{OK}, \texttt{FAIL}, \texttt{CHEATING}, \texttt{TIMEOUT}, or \texttt{ERROR}.

Our primary reported baselines use GPT-5.3-Codex in medium-effort and high-effort
modes under the evaluation configuration described above with a Zero-shot, query-only assessment. Each mode produces one
answer per problem.

Out of the full set of \textbf{2,284} benchmark problems, the medium-effort run
produced \textbf{743} answers that passed syntax checking and reached proof
execution, while the high-effort run produced \textbf{766}; the remaining
answers were excluded at the syntax-check stage because they did not compile as
valid OCaml/HOL Light tactic expressions.

Across the full benchmark, the medium-effort run solved \textbf{101 / 2,284}
problems (\textbf{4.4\%}) and the high-effort run solved \textbf{121 / 2,284}
(\textbf{5.3\%}), a net gain of \textbf{+20} proofs. Gains concentrate in the
\texttt{program\_state} category (+12 problems) and \texttt{generic} (+7 problems). We also host a \href{https://kings-crown.github.io/s2n-bignum-leaderboard/}{leader-board} for other research groups to make their own attempts on this problem set.

\begin{center}
\resizebox{\linewidth}{!}{
\begin{tabular}{l rrrrrr rrrrrr}
\toprule
& \multicolumn{6}{c}{\textbf{Medium Effort}} 
& \multicolumn{6}{c}{\textbf{High Effort}} \\
\cmidrule(lr){2-7} \cmidrule(lr){8-13}
\textbf{Category} 
  & \textbf{Total} & \textbf{Eval} & \textbf{OK} & \textbf{FAIL} 
  & \textbf{TO/ERR} & \textbf{OK/Tot}
  & \textbf{Total} & \textbf{Eval} & \textbf{OK} & \textbf{FAIL} 
  & \textbf{TO/ERR} & \textbf{OK/Tot} \\
\midrule
generic          & 562 & 330 & 59 & 239 & 32 & 10.5\% 
                 & 562 & 349 & 66 & 246 & 37 & 11.7\% \\
bit\_vector      & 311 & 142 & 26 & 109 &  7 &  8.4\% 
                 & 311 & 140 & 27 & 107 &  6 &  8.7\% \\
program\_state   & 552 & 235 & 16 & 211 &  8 &  2.9\% 
                 & 552 & 229 & 28 & 193 &  8 &  5.1\% \\
fc\_arm          & 437 &  33 &  0 &  33 &  0 &  0.0\% 
                 & 437 &  42 &  0 &  42 &  0 &  0.0\% \\
fc\_x86          & 422 &   3 &  0 &   3 &  0 &  0.0\% 
                 & 422 &   6 &  0 &   6 &  0 &  0.0\% \\
\midrule
Total            & 2,284 & 743 & 101 & 595 & 47 & 4.4\% 
                 & 2,284 & 766 & 121 & 594 & 51 & 5.3\% \\
\bottomrule
\end{tabular}
}
\end{center}
\noindent\small{TO/ERR combines \texttt{TIMEOUT} and 
\texttt{ERROR} verdicts. In aggregate, the medium-effort run 
produced 44 timeouts and 3 errors; the high-effort run 
produced 47 timeouts and 4 errors.}


\end{document}